\documentstyle[epsfig,11pt]{article}
\def\be{\begin{eqnarray}}
\def\ee{\end{eqnarray}}

\def\half{{\textstyle \frac{1}{2}}}

\def\roughly#1{\mathrel{\raise.3ex\hbox{$#1$\kern-.75em%
\lower1ex\hbox{$\sim$}}}}

\def\bfrho{{\mbox{\boldmath $\rho$}}}
\def\bfalpha{{\mbox{\boldmath $\alpha$}}}

\begin{document}

\renewcommand{\thefootnote}{\arabic{footnote}}
\setcounter{footnote}{0}

\vskip 0.4cm
\hfill {\bf KIAS-P00043}\\
$\mbox{}$\hfill {\bf FZJ-IKP(TH)-2000-16}\\

\hfill {\today}
\vskip 1cm

\begin{center}
{\LARGE\bf $\tilde\pi^0\rightarrow
\tilde\gamma\tilde\gamma$ in Dense QCD}

\date{\today}

\vskip 1cm
Maciej A. Nowak$^{a,b}$\footnote{E-mail: nowak@kiwi.if.uj.edu.pl },
Mannque Rho$^{a,c}$\footnote{E-mail: rho@spht.saclay.cea.fr},
Andreas Wirzba$^d$\footnote{E-mail: a.wirzba@fz-juelich.de}
and Ismail Zahed$^{a,e}$\footnote{E-mail: zahed@zahed.physics.sunysb.edu}

\end{center}

\vskip 0.5cm

\begin{center}

$^a$
{\it School of Physics, Korea Institute for Advanced Study, 
Seoul 130-012, Korea}

$^b$
{\it M. Smoluchowski Institute of Physics,
Jagellonian University, Cracow, Poland}

$^c$
{\it Service de Physique Th\'eorique, CE Saclay,
91191 Gif-sur-Yvette, France}

$^d${\it  FZ J\"ulich, Institut f\"ur Kernphysik, D-52425 J\"ulich, Germany}

$^e$
{\it Department of Physics and Astronomy,
SUNY-Stony-Brook, NY 11794, U.\,S.\,A.}

\end{center}

\vskip 0.5cm

\begin{abstract}
QCD superconductors in the color-flavor-locked (CFL) phase
support light excitations (generalized pions) in the form of
particle-particle or hole-hole excitations.
We analyze the generalized process
$\tilde\pi^0\rightarrow\tilde{\gamma}\tilde{\gamma}$ 
in the weak coupling limit and show that it is related
to the recently suggested Wess-Zumino-Witten (WZW) term. 
In dense QCD, the
radiative decay of the generalized pion is constrained
by geometry and vanishes at large density.

\end{abstract}
\newpage

\renewcommand{\thefootnote}{\#\arabic{footnote}}
\setcounter{footnote}{0}


{\bf 1.\,\,\,}
At high density, QCD exhibits a superconducting phase with
novel and nonperturbative phenomena~\cite{ALL,ALL98}. For
degenerate quark masses the QCD superconductor breaks color
and flavor spontaneously with the occurrence of Goldstone
modes. These modes were analyzed recently using effective 
Lagrangians~\cite{HRZ,GATTO,SONSTE} (zero size)
and bound state equations~\cite{RWZ,RSWZ} (finite size).

In the color-flavor-locked (CFL) phase 
the Goldstone modes are characterized by
a Wess-Zumino-Witten (WZW) term 
with a generalized axial-anomaly~\cite{HRZ}.
In the present letter we analyze the decay process 
$\tilde\pi^0\rightarrow\tilde\gamma\tilde\gamma$ of the
generalized pion $\tilde \pi^0$ 
into a pair of modified photons $\tilde \gamma$  using 
weak-coupling arguments at large quark chemical potential. After 
recalling some features of the QCD superconductor, we proceed to an 
analysis of the triangle anomaly in the CFL phase. We derive
an explicit result for the amplitude of 
$\tilde\pi^0\rightarrow\tilde\gamma\tilde\gamma$ in the limit of
large quark density. We show that our result agrees with the recently
suggested WZW term~\cite{HRZ}. The decay amplitude and the color-flavor
anomalies in the CFL phase are related at asymptotic densities.

\vskip 1.5cm
{\bf 2.\,\,\,}
In the QCD superconductor, the quarks are gapped. Their propagation in the
chiral limit is 
given in the Nambu-Gorkov formalism by~\cite{RWZ,PisarskiSuperfluid,PISARSKI}
\begin{eqnarray}
S_{11}(q)&=&\frac{1}{i}\left< \psi(q)\bar{\psi}(q)\right> =
\left[ \Lambda^+({\bf q}) \gamma^0\,
 \frac{q_0+q_{||}}{q_0^2-{\bf \epsilon}^2_q} \Lambda^{-}({\bf q}) 
+ \Lambda^-({\bf q}) \gamma^0\,
 \frac{q_0-q_{||}-2\mu}{q_0^2-\bar{{\bf \epsilon}}^2_q} \Lambda^{+}({\bf q}) 
\right]   \nonumber \\
S_{12}(q)&=&\frac{1}{i}\left< \psi(q)\bar{\psi}_C(q)\right> =
- \left[ \Lambda^+{\bf q}) {\bf M}^{\dagger}
 \frac{G^{*}(q)}{q_0^2-{\bf \epsilon}^2_q} \Lambda^{+}({\bf q}) 
+ \Lambda^-({\bf q}) {\bf M}^{\dagger}
 \frac{\bar{G^{*}}(q)}{q_0^2-\bar{{\bf \epsilon}}^2_q} \Lambda^{-}({\bf q}) 
\right] \nonumber \\ 
S_{21}(q)&=&\frac{1}{i}\left< \psi_C(q)\bar{\psi}(q)\right> =
\left[ \Lambda^-({\bf q}) {\bf M}
 \frac{G(q)}{q_0^2-{\bf \epsilon}^2_q} \Lambda^{-}({\bf q}) 
+ \Lambda^+({\bf q}) {\bf M}
 \frac{\bar{G}(q)}{q_0^2-\bar{{\bf \epsilon}}^2_q} \Lambda^{+}({\bf q}) 
\right]  \nonumber \\
S_{22}(q)&=&\frac{1}{i}\left< \psi_C(q)\bar{\psi}_C(q)\right> =
\left[ \Lambda^-({\bf q}) \gamma^0\,
 \frac{q_0-q_{||}}{q_0^2-{\bf \epsilon}^2_q} \Lambda^{+}({\bf q}) 
+ \Lambda^+({\bf q}) \gamma^0\,
 \frac{q_0+q_{||}+2\mu}{q_0^2-\bar{{\bf \epsilon}}^2_q}
 \Lambda^{-}({\bf q}) 
\right] \nonumber
\label{PropPart}
 \end{eqnarray}
with the momentum $q_{||}=(|{\bf q}|-\mu)$ measured parallel and 
relative to the Fermi momentum 
$|{\bf p}_F|=\mu$, the squared particle/hole energies 
${\bf {\epsilon}}_q^2 =q_{||}^2+{\bf M}^{\dagger}{\bf M}|G(q)|^2$ 
and the squared anti-particle/anti-hole energies
$\bar{{\bf{\epsilon}}}_q^2 =(q_{||} +2\mu)^2+{\bf M}^{\dagger}{\bf M}
|\bar{G}(q)|^2$. The operators $\Lambda^{\pm}({\bf q})=
\half(1\pm \bfalpha\cdot\hat{{\bf q}})$ are the
positive and negative energy projectors. In the CFL phase,
${\bf M}= \epsilon_f^a\epsilon_c^a\,\gamma_5= {\bf M}^\dagger$ is the locking
matrix in the color-flavor ($c$-$f$) sector with
$(\epsilon^a)^{bc}=\epsilon^{abc}$ the totally antisymmetric tensor.
To leading logarithm accuracy, the gap function $G(q)$ is 
solely a real-valued function of $q_{||}$ given by
\cite{PISARSKI,SON1,SchaferWilczek}
\begin{equation}
 G(x)=G_0\,\sin \left(\frac {\pi x}{2x_0}\right)
 = G_0 \sin \left(h_* x/\sqrt{3}\right)\; .
 \label{GapSolution}
\end{equation}
Here the logarithmic scales are defined as
$x=\ln(\Lambda_*/q_{||})$ and
$x_0={\rm ln}(\Lambda_*/G_0)$, with
$\Lambda_*=(4\Lambda_{\perp}^6/\pi m_E^5)$ in terms of the transversal cutoff 
$\Lambda_{\perp}=2\mu$ and the electric screening mass 
$m_E=\sqrt{\frac{N_F}{2 \pi^2}} g\mu$, where $g$ is the 
strong coupling constant
and $N_F$  the number of flavors.  Also, we have defined 
$h_* x_0/\sqrt{3}=\pi/2$, where $h_*=g/(\sqrt{6}\pi)$, 
and $G_0$ as
\begin{eqnarray}
 G_0 \approx \left(\frac{4 \Lambda_{\perp}^6}{\pi m_E^5}\right)\,
  e^{-\frac{\sqrt{3}\pi}{2h_\ast}} \; .
\label{Gzero}
\end{eqnarray}

The wave-functions of the pseudoscalar excitations
in the CFL phase have been discussed in~\cite{RWZ,RSWZ}.
In particular, in leading logarithm approximation, the pseudoscalar 
vertex operator of `isospin' $A$ for a pair of particles
or holes with momenta $P/2\pm p$ 
is given by
\begin{eqnarray}
  {\bf \Gamma}^A_{\pi}\,(p,P)= \frac {1}{F_T}\,\,
\left(\begin{array}{cc} 0 &  -i\gamma_5\,\Gamma_{PS}^\ast(p,P)\,
      \left({\bf M}^{A}\right)^\dagger 
                  \\
            i\gamma_5\,\Gamma_{PS} (p,P)\,{\bf M}^A 
  & 0\end{array}\right)
\label{PSvertex}
 \end{eqnarray}
with ${\bf M}^A={\bf M}^{a\alpha}\,( \tau^A)^{a\alpha}$
and ${\bf M}^{a\alpha}=\epsilon_f^a\,\epsilon_c^\alpha\,
\gamma_5$. Note that $\left({\bf M}^A\right)^\dagger = \epsilon_f^a
\epsilon_c^\alpha \gamma_5
({\tau^A}^\ast)^{a\alpha}= \epsilon_f^a
\epsilon_c^\alpha \gamma_5{(\tau^A)^{\alpha a}}$.
In the chiral limit, the pseudoscalar vertex reduces to the gap,
i.e. $\Gamma (p, 0) = G(p)$. Throughout, $F_T=\mu/\pi$ refers to the
temporal pion decay constant~\cite{HRZ,SONSTE,RWZ}. The CFL phase supports other collective
excitations as well~\cite{RSWZ}. They are not needed for the rest of 
our discussion.

\vskip 1.5cm
{\bf 3.\,\,\,}
In the CFL phase the ordinary photon is screened, and the 
gluons are either screened or higgsed. However, it was pointed out
in~\cite{CFL_ARW} that the CFL phase is transparent to a modified
or tilde photon,
\begin{equation}
\tilde{A}_\mu= A_\mu\, {\rm cos}\theta + H_\mu\,{\rm sin}\theta\; ,
\label{TILDE}
\end{equation}
where $\cos\theta=g/\sqrt{e^2+g^2}$, $A_\mu$ is the photon field
coupling to the charge matrix $e\,{\bf Q}_{em}$ of the quarks and
 $H_\mu$ is the gluon field for $U(1)_Y$ where $g\,{\bf Y}$ is the
color-hypercharge matrix~\footnote{Without loss of generality,
the representation of ${\bf Y}$ 
can be chosen to be identical to the one of ${\bf Q}_{em}$.}
and $\sin\theta=-e/\sqrt{e^2+g^2}$.
$\tilde{A}_\mu$ carries color-flavor and tags to the
charges of the Goldstone modes. 
The quark coupling to the tilde photon is in units of
$\tilde{e}=e\cos\theta$. As a result, the CFL phase is
characterized by generalized flavor-color anomalies~\cite{HRZ}.
In this section, we show  how
the triangle `anomaly' emerges from a direct calculation of the
$\tilde\pi^0\rightarrow\tilde\gamma\tilde\gamma$ in the leading
logarithm approximation. 

The contribution of (\ref{TILDE}) 
to the triangle graph is
\begin{eqnarray}
{\cal T}^{A}_{\mu\nu}(K_1,K_2)&=&-\,
\int\,\frac{d^4q}{(2\pi)^4}\,
{\rm Tr}\left(\,i{\bf \Gamma}_{\pi}^A (q,P)\,
i{\bf S}(q-\frac P2 )\,i\widetilde{\bf Q}_\mu
i{\bf S}(q+\frac Q2)\,i\widetilde{\bf Q}_\nu
\,i{\bf S} (q+\frac P2)\right)\,\, \nonumber \\
&&\qquad\qquad\qquad\quad 
 + \left[ ( K_1 \leftrightarrow K_2)  \,\,\,\,{\rm and}\,\,\,\,\, (\mu 
 \leftrightarrow \nu) \right]
\label{DEC11}
\end{eqnarray}
with $P=K_1+K_2$ and $Q=K_2-K_1$. The spin-color-flavor vertex due to
(\ref{TILDE}) reads
\begin{eqnarray}
  \widetilde{\bf Q}_\mu=\tilde{e}\,\bfrho_3\,\gamma_\mu\,
{\rm diag}(\widetilde{\bf Q}, {\widetilde{\bf Q}}^T)
=\tilde{e}\,\gamma_\mu\,\bfrho_0\,\widetilde{\bf Q}=
\tilde{e}\,\gamma_\mu\,\bfrho_0\,\,
\left( {\bf Q}_{em}\otimes {\bf 1}_c - {\bf 1}_f\otimes{\bf Y}\right)\;.
 \label{CHARGE1}
\end{eqnarray}
Here $\bfrho_3$ and $\bfrho_0$ are Pauli matrices acting on the Nambu-Gorkov
indices.
We assume that the indices $\mu$ and $\nu$ are eventually contracted
with the polarizations ${\cal E}^\mu_T$ and ${\cal E}^\nu_T$
(space-like).
The result for (\ref{DEC11}) after inserting (\ref{PropPart}) and
(\ref{PSvertex}) and 
lengthy algebra  is
\begin{eqnarray}
{\cal T}^{A}_{\mu\nu}(K_1,K_2)&= &2 i\frac{{\tilde e}^2}{F_T}
\int\,\frac{d^4q}{(2\pi)^4}\,\Gamma_{PS}\, (p, P)\,\,
\left[ \Sigma_{\mu \nu}^{A \,\,(0)}( P_i, {\bf M}) + 
\Sigma_{\mu \nu}^{A\,\,(1)}(P_i,
{\bf M}) + {\cal O}(1/\mu^{3})\right] \nonumber \\
&& +\left(
   {P_{i}}_+ \leftrightarrow {P_{i}}_- \ \mbox{or} \
{P_i}_{-}-2\mu   \leftrightarrow {P_{i}}_{+}+2\mu \ \mbox{and}\  
     \Lambda^+ \leftrightarrow \Lambda^-\;, {\bf M}^{(A)} \leftrightarrow
   {{\bf M}^{(A)}}^\dagger\right) \nonumber\\
\label{MAIN}
\end{eqnarray}
with $P_1\equiv q-P/2$, $P_2\equiv q+Q/2$, $P_3\equiv q+P/2$,
${P_i}_\pm \equiv {P_i}_0 \pm {P_i}_{||}$.  
The explicit form of $\Sigma_{\mu \nu}^{A\, (0)}$ is
\begin{eqnarray}
\Sigma_{\mu \nu}^{A\,\,(0)}&=& \left \{
    {\rm Tr}_{cf}
     [{\bf M}^{A} \,\widetilde{\bf Q}\,\widetilde{\bf Q}\, {\bf M}^\dagger]
                  \ {P_1}_+ {P_2}_+ G(P_3) \right. \nonumber\\
&& 
  \mbox{}-{\rm Tr}_{cf}
     [{\bf M}^{A}\, \widetilde{\bf Q}\,{\bf M}^\dagger\,\widetilde{\bf Q}]
                  \  {P_1}_+ G(P_2) {P_3}_{-} \nonumber\\
&&
  \mbox{}+{\rm Tr}_{cf}
     [ {\bf M}^{A}\,{\bf M}^\dagger\, \widetilde{\bf Q}\,\widetilde{\bf Q}] 
             \  G(P_1) {P_2}_- {P_3}_{-}\nonumber\\
&&
 \mbox{}\left. -{\rm Tr}_{cf}
         [ {\bf M}^{A}\,{\bf M}^\dagger\, \widetilde{\bf Q}\, {\bf M} 
    \, \widetilde{\bf Q}\,{\bf M}^\dagger]\ G(P_1) G(P_2) G(P_3) \right\}
\nonumber\\
&& \mbox{} \times
   \frac{1}{\Delta(P_1) \Delta(P_2) \Delta(P_3)}
{\rm Tr}[ \gamma_5 \Lambda^+({\bf P}_1) \gamma_\mu 
      \Lambda^{-}({\bf P}_2)
             \gamma_\nu \Lambda^{+}({\bf P}_3)] \;. 
\label{ANO1}
\end{eqnarray}
The explicit form of $\Sigma_{\mu \nu}^{A \,\,(1)}$ is
\begin{eqnarray}
\Sigma_{\mu \nu}^{A\,\,(1)} &=&  \left\{
  \frac{1}{\Delta(P_1)\bar{\Delta}(P_2)\Delta (P_3)} 
  {\rm Tr}_{cf}
     [{\bf M}^{A} \,\widetilde{\bf Q}\,\widetilde{\bf Q}\, {\bf M}^\dagger]
                  \  {P_1}_{+} ({P_2}_{-}-2\mu) G(P_3) \right. 
\nonumber \\
&& \mbox{} - \frac{1}{\bar{\Delta}(P_1)\Delta(P_2)\bar{\Delta} (P_3)}
  {\rm Tr}_{cf}
     [{\bf M}^{A}\, \widetilde{\bf Q}\,{\bf M}^\dagger\,\widetilde{\bf Q}]
                  \  ({P_1}_{-}-2\mu) G(P_2) ({P_3}_{+}+2\mu) \nonumber\\
&& \mbox{} - \frac{1}{{\Delta}(P_1)\bar{\Delta}(P_2){\Delta}(P_3)}
{\rm Tr}_{cf}
     [{\bf M}^{A}\, \widetilde{\bf Q}\,{\bf M}^\dagger\,\widetilde{\bf Q}]
                  \  {P_1}_{+} \bar{G}(P_2) {P_3}_{-} \nonumber\\
&& \mbox{} + \frac{1}{\Delta(P_1)\bar{\Delta}(P_2)\Delta (P_3)}
   {\rm Tr}_{cf}
     [ {\bf M}^{A}\,{\bf M}^\dagger\, \widetilde{\bf Q}\,\widetilde{\bf Q}] 
             \  G(P_1) ({P_2}_{+}+2\mu) {P_3}_{-}\nonumber\\
&&\left. \mbox{} - \frac{1}{\Delta(P_1)\bar{\Delta}(P_2)\Delta (P_3)}
   {\rm Tr}_{cf}
         [ {\bf M}^{A}\,{\bf M}^\dagger\, \widetilde{\bf Q}\, {\bf M} 
    \, \widetilde{\bf Q}\,{\bf M}^\dagger]
\ G(P_1) \bar{G}(P_2) G(P_3)\,\, \right\}
\nonumber\\
&& \mbox{} \times
  {\rm Tr}[ \gamma_5 \Lambda^+({\bf P}_1) \gamma_\mu 
      \Lambda^{+}({\bf P}_2)
             \gamma_\nu \Lambda^{+}({\bf P}_3) ]\;.
 \label{ANO2}
\end{eqnarray}
We have defined $\Delta(P)=P^2 -{\bf \epsilon}^2_{P}$,
$\overline{\Delta}(P)=P^2-\overline{\epsilon}^2_P$, and made
use of the fact that in the generalized pion CM frame 
${\bf P}_1={\bf P}_3$, thereby reducing the spin
contributions~\footnote{Because of 
$ \gamma_5\Lambda^\pm ({\bf P}_1) = \gamma_5\Lambda^\pm ({\bf P}_3)
=\Lambda^\pm ({\bf P}_3)\gamma_5$, 
the left-out spin contributions project to zero  after a
cyclic permutation under the Dirac trace.}.

\begin{figure}
\centerline{\epsfig{file=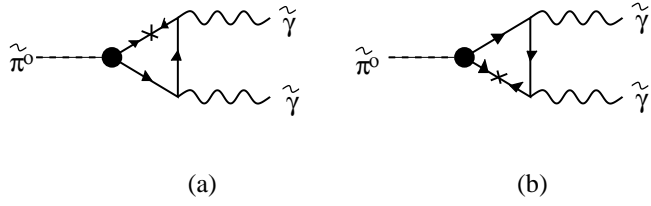,height=1in}}
\caption{Leading contributions to
$\tilde{\pi}^0\rightarrow\tilde\gamma\tilde\gamma$ in the CFL phase.}
\end{figure}

The various contributions in $\Sigma^{\,\,(0)}$ 
arise from particles and holes. Naively,  $\Sigma^{(0)}$
generates a term  of order $\mu^2$ as it receives contribution from
the full Fermi surface. We show below that this contribution
is identically zero after integration. The various contributions
to $\Sigma^{(1)}$ involve at least an antiparticle.
Since $\overline{\Delta}\approx -4\mu^2$, the overall contribution
is of order $\mu^0$ after integrating over the Fermi 
surface~\footnote{The terms
linear and quadratic in $\mu$ cancel when (\ref{ANO2}) is inserted
into (\ref{MAIN}).}. It can be shown 
that the non-vanishing contributions arising from the various
terms in $\Sigma^{(1)}$ are those displayed in Fig.~1, where the cross
refers to the pair-condensate insertion. They correspond to 
the first and fourth line  of
(\ref{ANO2}) (and first and third line of (\ref{ANO1})) when inserted
into (\ref{MAIN}), 
and they
saturate 
the photo-decay of the pion in dense QCD. Finally, we note the different
structure in the Lorentz projectors in (\ref{ANO1}) and (\ref{ANO2}). 
To the order considered, we use ${\bf M}^{\dagger}{\bf M} =1$
in $\Delta$ and $\bar{\Delta}$~\footnote{Under the assumption that $U_A(1)$
is broken by the anomaly.}.
The overall factor of 2 in front of (\ref{MAIN}) comes from the triangle
graph with crossed tilde-photon legs. 
The various color-flavor traces yield equal contributions modulo
overall signs. Specifically, 
\begin{eqnarray}
&&{\rm Tr}_{cf}\left( {\bf M^A}\tilde{\bf Q}\tilde{\bf Q}{\bf M}^\dagger\right) =
-2N_f\,{\rm Tr} \left( \tau^A\,\tilde{\bf Q}_*^2\right)\nonumber\\
&&{\rm Tr}_{cf}\left( {\bf M^A}\tilde{\bf Q}{\bf M}^\dagger\tilde{\bf Q}\right) =
+2N_f\,{\rm Tr} \left( \tau^A\,\tilde{\bf Q}_*^2\right)\nonumber\\
&&{\rm Tr}_{cf}\left( {\bf M^A}{\bf M}^\dagger\tilde{\bf Q}\tilde{\bf Q}\right) =
-2N_f\,{\rm Tr} \left( \tau^A\,\tilde{\bf Q}_*^2\right)\nonumber\\
&&{\rm Tr}_{cf}\left( {\bf M^A}{\bf M}^\dagger\tilde{\bf Q}{\bf M}\tilde{\bf Q}{\bf M}^\dagger\right) =
+2N_f\,{\rm Tr} \left( \tau^A\,\tilde{\bf Q}_*^2\right)\,\,,
\label{TRACES}
\end{eqnarray}
where we have defined 
\begin{eqnarray}
  \widetilde{\bf Q}_*= \frac 13\,\,
  \left(\begin{array}{ccc} 2& 0 & 0\\0 & -1& 0\\ 0 &0& -1
  \end{array}\right)\; .
 \label{CHARGE}
\end{eqnarray}
The overall factor of 2 in (\ref{TRACES}) stems from an equal contribution
due to either flavor or color, since in the CFL phase $N_f=N_c$.

Inserting (\ref{TRACES}) into (\ref{ANO1}) and performing the energy 
integration over the Fermi surface yield 
\begin{eqnarray}
{\cal T}^{A\,\,(0)}_{\mu\nu} (K_1, K_2) = {\cal R}_0\,\, {\cal I}_0\,\, 
\frac{\tilde{e}^2}{2\pi^2 F_T}\,
{\rm Tr} \left( \tau^A\,\widetilde{\bf
    Q}_*^2\right)\,\epsilon_{\mu\nu\alpha\beta}\,K_1^\alpha\,K_2^\beta
 \; ,
\label{ANO1a}
\end{eqnarray}
where ${\cal R}_0$ is an overall constant  and ${\cal I}_0$ is an integral of the form
\begin{eqnarray}
\int d^4q \frac{G^2 (q)}{\Delta^3 (q)}  \sim \frac{\mu^2}{G_0^2} \; .
\label{int0}
\end{eqnarray}
The $\mu^2$ on the r.h.s. of (\ref{int0}) arises from integrating over
the full Fermi surface. This contribution 
implies a decay rate that increases substantially in dense matter.
Careful algebra,  however, shows that after integration over the
Fermi surface, the sum of the various contributions to $\Sigma^{(0)}$ yields 
${\cal R}_0= 0$. The decay rate is driven by $\Sigma^{(1)}$ to leading
order, as we now show.

To facilitate the calculation of the various contributions to
$\Sigma^{(1)}$, we choose the CM kinematics with $P_1=(q^0-M/2, \vec{q})$,
$P_2=(q^0+Q^0/2, \vec{q}+\vec{Q}/2)$, $P_3 =(q^0 +M/2, \vec{q})$.
After some algebra we obtain 
\begin{eqnarray}
{\cal T}^{A\,\,(1)}_{\mu\nu} (K_1, K_2) =  2  \frac{\tilde{e}^2}{F_T}
\frac{N_f}{3} \,\,\,\frac{{\cal{I}}_1}{\mu^2} \,\,\,
{\rm Tr}\left( {\tau^A}\,\widetilde{\bf
    Q}_*^2\right)\,\epsilon_{\mu\nu\alpha\beta}\,K_1^\alpha\,K_2^\beta
 \; .
\label{ANO1b}
\end{eqnarray}
Note that the $1/3$ results from an angular integration over the
3-momenta through the Fermi surface, while
\begin{eqnarray} 
{\cal I}_1=\int \frac{d^4q}{(2\pi)^4}\frac{G^2(q)}{\Delta^2(q)} =\frac{i}{8} F^2_T=
\frac{i\,\mu^2}{8\pi^2} 
\end{eqnarray} 
is proportional to the square of the temporal pion decay constant~\cite{SONSTE,RWZ}. 
The presence of the antiparticle yields a $\mu$-independent 
contribution~\footnote{The same observation applies to most of the mixing processes
discussed in~\cite{RSWZ} and found to vanish to leading order.}.
Note that the various contributions involving the
antiparticle gap $\bar{G}$ in $\Sigma^{(1)}$ 
contribute
zero after integration. The same is true for the terms with
any pair-condensate
insertion on the fermion leg in between the two photon vertices.
Hence,
\begin{eqnarray}
{\cal T}^{A}_{\mu\nu} (K_1, K_2) =   i \frac{\tilde{e}^2}{F_T}
\frac{1}{4\pi^2}
{\rm Tr}\left({\tau^A}\,\widetilde{\bf
    Q}_*^2\right)\,\epsilon_{\mu\nu\alpha\beta}\,K_1^\alpha\,K_2^\beta
\label{RESULTALL}
\end{eqnarray}
to leading order in the density. We have checked that the
possible vertex corrections to (\ref{PSvertex}) are subleading
in the decay rate. Because of the $F_T$-dependence, 
eq. (\ref{RESULTALL}) suggests that the radiative
decay of the generalized pion vanishes as $1/\mu$ in dense
matter.

This result, based on our explicit calculation in the CFL
phase to leading order, is in agreement with the result
suggested by the generalized WZW term~\cite{HRZ},
\begin{equation}
 \partial^\mu\,{\bf A}^3_{\mu} =-\frac{\tilde{e}^2}{96\pi^2}\,
 \epsilon_{\mu\nu\rho\sigma}\,
 \tilde{F}^{\mu\nu}\,\tilde{F}^{\rho\sigma}\,\,
\label{ANOMALYCF}
\end{equation}
with $\tilde{F}^{\mu\nu}$ the field strength associated to 
(\ref{TILDE})~\footnote{The normalization of the
$\pi^0\gamma\gamma$ amplitude is 8 times bigger  than the
normalization of the
corresponding  effective Lagrangian or of the pertinent anomaly
(\ref{ANOMALYCF}).  
Furthermore,  ${\rm Tr}\left({\tau^3}\,\widetilde{\bf
    Q}_*^2\right)= 1/3$.}.
The finite-size of the generalized pion does not upset the geometrical
normalization of the anomaly in the generalized WZW form~\cite{HRZ}.
A direct comparison to the $\mu=0$ radiative decay of the usual pion in QCD, 
shows that (\ref{ANOMALYCF}) is off by a factor of
$1/3$.  As originally
noted in~\cite{HRZ}, the color-flavor anomaly in the CFL phase is no
longer multiple of $N_c$ due to the color-flavor locking in the triangle
graph, hence a factor of $1/3$.
This point is particularly explicit in the various traces in (\ref{TRACES}).

\vskip 1.5cm
{\bf 4.\,\,\,}
In the CFL phase ordinary photons are screened, and perturbative
gluons are either screened or higgsed. The anomalous decay of the generalized pions occurs via the tilde-photon.
We have provided a direct calculation of the radiative decay $\tilde\pi^0\rightarrow\tilde\gamma\tilde\gamma$
in this phase using weak-coupling arguments. 
To leading logarithm accuracy our result is exact and in agreement with the normalization suggested by the
generalized WZW term given in~\cite{HRZ}. Much like in the vacuum, the radiative decay  of the generalized pions 
is dictated by geometry in leading order, and vanishes at asymptotic densities. Clearly our analysis extends to
other anomalous as well as nonanomalous processes, and should prove useful for the analysis of emission rates
in dense QCD.

\vskip 2cm
{\bf Acknowledgments:}
\vskip .5cm
M.A. Nowak, M. Rho and I. Zahed  thank KIAS for hospitality 
during the completion of this work.
This work was supported in part by the US DOE grant DE-FG02-88ER40388
and by the Polish Government Project (KBN)  2P03B 00814.

\end{document}